\documentclass[twocolumn,showpacs,amsmath,amssymb]{revtex4}

\usepackage{graphicx}
\usepackage{dcolumn}
\usepackage{bm}

\begin{document}

\title{Measurement of coherent
charge transfer in an adiabatic Cooper pair pump}

\author{Rosario Fazio}
\affiliation{NEST-INFM \& Scuola Normale Superiore, 56126 Pisa, Italy}

\author{F.W.J. Hekking}
\affiliation{Laboratoire de Physique et Mod\'elisation des Milieux Condens\'es\\
Magist\`ere-CNRS, B.P. 166, 38042 Grenoble cedex 9, France}

\author{J.P. Pekola}
\affiliation{Low Temperature Laboratory, Helsinki University of Technology, P.O. Box
2200, FIN-02015 HUT, Finland}

\begin{abstract}
We study adiabatic charge transfer in a superconducting Cooper pair pump, focusing on
the influence of current measurement on coherence. We investigate the limit where the
Josephson coupling energy $E_J$ between the various parts of the system is small
compared to the Coulomb charging energy $E_C$. In this case the charge transferred in
a pumping cycle $Q_P \sim 2e$, the charge of one Cooper pair: the main contribution is
due to incoherent Cooper pair tunneling. We are particularly interested in the quantum
correction to $Q_P$, which is due to coherent tunneling of pairs across the pump and
which depends on the superconducting phase difference $\varphi _0$ between the
electrodes: $1-Q_P/(2e) \sim (E_J/E_C) \cos \varphi _0$. A measurement of $Q_P$ tends
to destroy the phase coherence. We first study an arbitrary measuring circuit and then
specific examples and show that coherent Cooper pair transfer can in principle be
detected using an inductively shunted ammeter.
\end{abstract}

\pacs{74.50.+r, 74.78.Na, 73.23.Hk}

\maketitle

\section{Introduction}
\label{intro}

It is well-known that electrons can be transferred through a mesoscopic device by
means of adiabatic changes of system parameters like, {\em e.g.}, externally applied
electric or magnetic fields. Since the original work of Thouless~\cite{thouless},
several theoretical proposals concerning possible physical realizations of this
phenomenon, usually referred to as parametric pumping, have been put
forward~\cite{moskalets}. Depending on the physical mechanism employed, these
proposals can be divided into two classes.

In open systems, {\em i.e.}, systems consisting of several conducting parts, connected
to each other by highly transmissive barriers, parametric pumping can be achieved
through a periodic modulation of phase of the scattering matrix associated to the
device~\cite{brouwer}. In these proposals, electronic phase coherence plays a
fundamental role and charge is transferred coherently through the entire system. The
amount of charge that is transferred per period of the modulation is in general a
fraction of the electronic charge $e$, which depends on the modulation path. In open
devices, electron-electron interactions are weak, and lead to (small) dephasing
corrections to the noninteracting result for charge transfer.

In the opposite limit of closed systems, {\em i.e.}, several metallic islands
connected to each other by ultrasmall tunnel junctions, parametric pumping of charge
is achieved by periodic modulation of the so-called Coulomb blockade~\cite{houches}.
In these proposals, the presence of strong Coulomb repulsion between electrons is
essential. It leads to the quantization of charge on the islands in units of $e$. A
periodic modulation of externally applied gate-voltages leads to a periodic lifting of
Coulomb blockade, which enables the transfer of exactly one electron per period
through the device. The presence of phase coherence in this limit only leads to small
corrections to the classical result, through coherent higher order charge transfer
processes known as (in)elastic co-tunneling~\cite{averinnazarov}.

Experimental evidence for parametric charge pumping in normal metallic systems has
been found in both limits of open~\cite{marcus} and closed~\cite{pothier,kouwenhoven}
systems.

The present paper is devoted to parametric pumping in a superconducting
system~\cite{geerligs}. It consists of superconducting islands, connected to each
other by small tunnel junctions, and it is operated in the Coulomb blockade regime.
From this point of view the system is closed, and charge transfer is mainly classical,
{\em i.e.} quantized in units of Cooper pair charge $2e$ per period of the modulation.
From the other hand phase is defined naturally in a superconducting system, and due to
the Josephson coupling between the various parts of the system  phase coherence tends
to be maintained throughout the device. In this sense, a superconducting pump behaves
as an open system, and the total charge transfer will be characterized by significant
phase-coherent corrections to the classical result~\cite{prb99}.

We will be particularly interested in the influence of the measurement apparatus of
the amount of transferred charge on the phase-coherent properties of a superconducting
charge pump. Measurement on a quantum coherent system is achieved by connecting it to
the environment provided by the measuring device. The environment in turn can be
modeled as a collection of harmonic oscillators. Depending on the environment, the
quantum coherence is typically lost in a time which we can call the dephasing or the
decoherence time once properly defined. We focus here on quantum coherent systems
formed of Josephson tunnel junctions, and their electromagnetic environment. These
systems \cite{nakamura,lukens,mooij,vion}, or we might already call some of them
devices to perform quantum logic operations (see review in \cite{makhlin}), hold great
promise in quantum computing because of their potential in scalability.

In a superconducting system the phase difference across a junction follows the
well-known AC and DC Josephson relations. The theory of phase fluctuations and
environment is well established for electrical circuits including small tunnel
junctions \cite{ingnaz,schonzaikin,martinis}. What we do here specifically is the
analysis of the back-action of the measurement of electrical current on a circuit
consisting of small Josephson junctions. In particular, we show that unlike in a
standard dissipative ammeter, the phase diffusion is limited in a measurement
performed by an inductively shunted measuring circuit. Measurement of tiny currents
provides a read-out of Josephson junction quantum bits, like the one in the so-called
"quantronium" experiment \cite{vion}, or in general when reading out the persistent
current in a flux quantum bit (an RF SQUID loop) \cite{lukens,mooij}. As a specific
example we focus on a measurement of current in a double island adiabatic Cooper pair
pump (CPP), but our conclusions concerning the effect of an inductance limited phase
diffusion measurement can easily be generalized to other Josephson junction circuits
with straightforward modifications, which account for the different topology of the
circuit to be measured.

\section{The model}

\begin{figure}
\includegraphics[width=8cm]{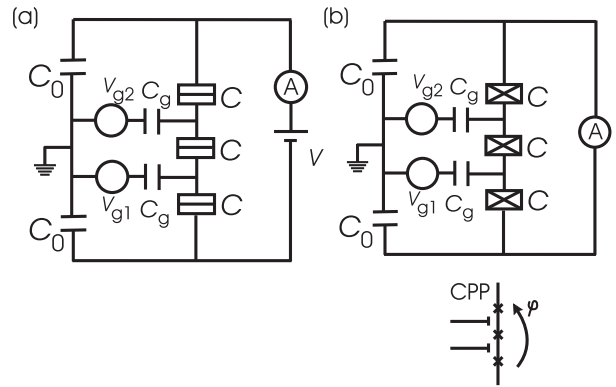}
\caption{\label{device}Double island (a) electron and (b) Cooper pair pump. A
symmetric pump consists of junctions with capacitance $C$ and Josephson coupling
energy $E_{J}$. Gate capacitance is $C_{g}$ for both islands, and the gate voltages
are $V_{g1}$ and $V_{g2}$ for the first and the second island, respectively. Electron
pump can be operated with either zero or non-zero bias voltage $V$ across; Cooper pair
pump has zero bias voltage but it can be phase biased by $\varphi$. In (b) we also
show the shorthand symbol used in the later figures for a CPP with phase $\varphi$
across.}
\end{figure}

An electron pump is a reversible device which provides quantized transport of
electrical charge upon cyclic operation of gates connected to it. The simplest variant
is perhaps a double island pump (see Fig.~\ref{device}a) in which tiny metallic
(non-superconducting) or semiconducting grains are tunnel coupled to each other and to
the surrounding electrodes, and capacitively coupled to two
gates~\cite{pothier,kouwenhoven}. The islands are so small that, due to their tiny
capacitance, equilibrium charge configurations on them are determined solely by
electrostatics on the single electron level characterized by the scale of charging
energy $E_{C}$. Applying $\pi/2$ phase shifted harmonic voltages to the two gates at a
frequency $f$, which is so low that it allows the system to follow the ground state
configuration of charging energy at each phase of operation, current through the pump
equals $I_P=ef$, based on transport of one electron per cycle, on a precision level of
a few per cent. A more accurate but more complicated device can be built by increasing
the number of islands and adjacent gates in the pump: this approach has been taken to
meet with requirements in metrology~\cite{keller}.

If all the electrodes and islands are superconducting (see Fig.~\ref{device}b), charge
can be transported not only as single electrons, or rather quasiparticles with single
electron charge in this case, but also as Cooper pairs~\cite{geerligs,prb99}. Ideally,
in the absence of any bias voltage and at low temperatures $T$ such that $k_BT \ll E_C
\ll \Delta$, where $\Delta$ is the superconducting gap and $E_C = 2e^2/C$ ($C$ is the
junction capacitance), the system is in the Coulomb blockade regime, and only Cooper
pairs are transferred. In this regime, the device can be referred to as a CPP,
described by the Hamiltonian
\begin{equation}
\hat{H}=\hat{H}_{C} + \hat{H}_{J} \;, \label{ham}
\end{equation}
where
\begin{eqnarray}
\hat{H}_{C} = \frac{2}{3}E_C \left[(\hat{n}_1 - n_{x1})^2 + (\hat{n}_2 -
n_{x2})^2 \right. \nonumber \\
\left. + (\hat{n}_1 - n_{x1})(\hat{n}_2 - n_{x2}) \right] \label{HC}
\end{eqnarray}
is the charging energy of the islands, coupled capacitively to each other as well as
to two gates $1$ and $2$. The operator $\hat{n}_{1(2)}$ denotes the number of excess
Cooper pairs on island 1(2) and $n_{x1(x2)}$ is the gate-charge (in units of $2e$)
which can be tuned by varying the gate voltage $V_{g1(g2)}$. The Josephson
contribution is given by
\begin{eqnarray}
\label{HJ} \hat{H}_{J} = -E_J\left[\cos (\varphi/3+\hat{\phi}_1)
+ \cos (\varphi/3 + \hat{\phi} _2 - \hat{\phi}_1) \right. \nonumber \\
\left.  + \cos (\varphi/3 - \hat{\phi}_2)\right] .
\end{eqnarray}
Here $E_{J}$ is the Josephson coupling energy of each junction (assumed equal for the
three junctions). Throughout this paper we will work in the limit $E_J \ll E_C$ where
charging effects dominate. The phase operator $\hat{\phi}_{1(2)}$ is conjugate to the
number operator $\hat{n}_{1(2)}$. The phase bias $\varphi$ is the phase difference
across the entire pump. As long as one can ignore any dynamical environment, $\varphi$
can be considered as a classical variable that can be set to a constant offset value
$\varphi = \varphi_0$, for instance, by applying a magnetic flux through the circuit
of Fig.~\ref{device}b.

\begin{figure}
\includegraphics[width=6cm]{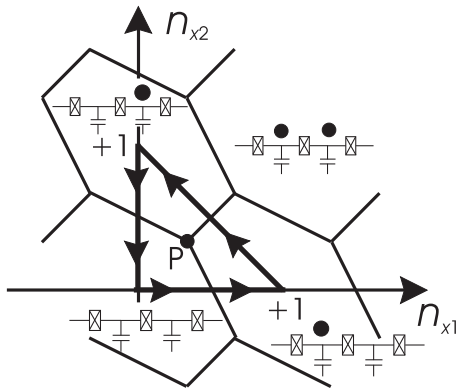}
\caption{\label{gating}Triangular gating sequence in the $n_{x1},n_{x2}$ plane for the
pumping cycle as discussed in the text. Hexagons correspond to regions in which the --
schematically indicated -- classical charge configurations are stable.}
\end{figure}

We are interested in the charge transferred through the device upon a periodic
modulation of the applied gate-charges $n_{x1} (t)$ and $n_{x2} (t)$. Specifically, we
assume the gate modulation to be adiabatically slow, with a frequency $f \ll
E_J^2/\hbar E_C$. Under this condition the system remains in the ground state
throughout the modulation if the temperature is low, $k_BT \ll E_J$. During one period
of the modulation, the two-component vector $\vec{n}_x = (n_{x1}, n_{x2})$ describes a
closed path in the corresponding two-dimensional plane, see, {\em e.g.},
Fig.~\ref{gating}. It can be shown~\cite{prb99} that the transferred charge in the
ground state during one period of the modulation is given by a contour integral along
the closed path followed by $\vec{n}_x$,
\begin{equation}
Q_P = 2  \hbar \Im \mbox{m} \left[ \sum \limits _{n \ne 0} \oint
\frac{(\hat{I})_{0n}}{E_0 - E_n} \langle n| \partial _{\vec{n}_x}0 \rangle \cdot
d\vec{n}_x \right] . \label{pumpchar}
\end{equation}
Here, we introduced the instantaneous eigenstates $|m \rangle $ and energies $E_m$ of
the corresponding time-dependent Hamiltonian~(\ref{ham}), such that $\hat{H}(t) |m
(t)\rangle = E_m(t) |m (t)\rangle$. The matrix element $(\hat{I})_{0n}$ corresponds to
the current operator of one of the junctions. For the leftmost junction for instance
we have
\begin{equation}
\hat{I} \equiv \hat{I}_l = I_J \sin(\hat{\phi}_1 + \varphi/3) , \label{lcur}
\end{equation}
where $I_J = 2e E_J /\hbar$ is the Josephson critical current. The matrix element is
taken between the instantaneous ground state $|0\rangle$ and the instantaneous excited
state $|n\rangle$. An outline of the derivation of~(\ref{pumpchar}) is given in
Appendix~A.

In the limit $E_J/E_C \to 0$, to leading order tunneling events involving more than
one junction can be ignored. As a result, a periodic gate modulation will lead to a
transfer of a charge $Q_{P}= 2e$ per cycle through the pump, independent of the bias
phase $\varphi_0$, see Appendix~\ref{appincoh}. This is the limit of incoherent Cooper
pair tunneling. Experimentally, incoherent Cooper pair tunneling is best realized by a
sufficiently dissipative environment which completely randomizes $\varphi_0$, {\em
e.g.}, by fabricating on-chip resistors near the pump, with resistances of the order
or larger than the resistance quantum for Cooper pairs: $R_K/4=h/(2e)^2 \simeq 6.45$
k$\Omega$~\cite{zorin}.

Deviations from the classical result are related to higher order tunneling processes.
In general, pumps in the superconducting state are less accurate than their normal
state counterparts and one fundamental reason is the coherence of the superconducting
wave function. In this paper we are interested in the regime where coherent Cooper pair
tunneling leads to an important correction to the incoherent charge transfer. It
provides an interesting, still unobserved quantum mechanical interference correction
to the pumped charge (or pumped current in continuous cycling) \cite{prb99}:
\begin{equation} \label{cohcor}
Q_{P}/2e\simeq 1-9(E_{J}/E_{C})\cos \varphi_0 .
\end{equation}
This result is obtained for a triangular gating, as shown in Fig.~\ref{gating}; some
details of the derivation are given in Appendix~\ref{appcoh}. Although the result is
perturbative in $E_{J}/E_{C}$, the coherent interference term may be appreciable. For
example for $E_{J}/E_{C}=0.1$, still rather well within the perturbative regime, the
variation of the pumped current $I_{P} \equiv Q_{P}f $, is 1.8 times larger than the
magnitude of the incoherent current $I_\mathrm{inc}\equiv 2ef$. The coherent
contribution to the current can be tuned by adjusting $\varphi_0$. By assuming a
realistic operating frequency of $f=100$ MHz, $2ef$ equals 32 pA, and this would
produce variation in the interference term by 58 pA. Detecting this phase coherent
current would not only be fundamentally interesting but it would also allow one to
measure dephasing in a superconducting multi-junction circuit coupled to a measuring
device.

We finally note that, in addition to the current $I_P = Q_P f$ discussed so far, a
direct Josephson current $I_\mathrm{jos}$ exists in response to the phase bias
$\varphi _0$. The Josephson current is maximum, $\sim I_J$, at the triple point P in
Fig.~\ref{gating}, where the three charge configurations $|0,0\rangle$, $|1,0\rangle$
and $|0,1\rangle$ are degenerate. The gating sequence of interest here is away from P:
at any given point along the cycle at most two out of the three aforementioned
relevant charge configurations are degenerate. At such a degeneracy, the Josephson
current is smaller, $\sim I_J (E_J/E_C)$. The width of the degeneracy is finite,
$\propto E_J/E_C$. Hence only a fraction of the total cycle contributes to the
Josephson current and $I_\mathrm{jos} \propto I_J (E_J/E_C)^2 \sin \varphi _0$. Thus
$I_\mathrm{jos}$ can be distinguished from $I_P$, as it depends differently on
$\varphi _0$ and is independent of the frequency $f$.

\section{Coupling to a measurement circuit}
\label{coupling}

\begin{figure}
\includegraphics[width=6cm]{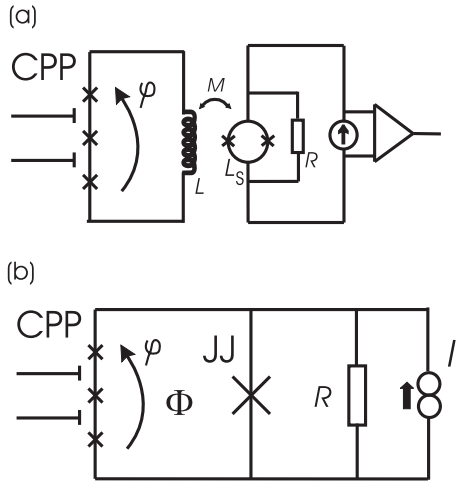}
\caption{\label{measure}Two specific examples of inductive measuring circuits for
current. In (a) the SQUID ammeter is inductively coupled to the measured circuit. In
(b) current is measured by determining the escape probability from the zero voltage
state of a Josephson junction. The inductance is provided by the measuring junction.
The phase $\varphi$ can be adjusted by, {\em e.g.}, applying a flux $\Phi$ through the
loop of the CPP and the measuring circuit.}
\end{figure}

We propose two strategies to measure the pumped charge as illustrated in
Fig.~\ref{measure}. The first variant, in Fig.~\ref{measure}a, is a generic
inductively coupled measurement by a DC SQUID (input coil $L$, mutual inductance $M$,
shunt resistance $R$). We call this circuit {\bf A} in the following. The second
circuit, in Fig.~\ref{measure}b, represents a directly coupled measurement of current
by a Josephson junction. In this configuration changes in current of the measured
circuit are detected as changes of lifetime of a meta-stable zero voltage
(supercurrent) state of the measuring current biased junction \cite{vion,balestro}.
This measurement circuit will be called {\bf B} in what follows. The measuring
junction is biased with a current $I$; its capacitance $C_{J}$ is the capacitance due
to its typically planar geometry. When biased near the critical current, $I \alt
I_{c}$, the junction can be characterized by the Josephson inductance
$L_{J}=\sqrt{2}\frac{\hbar}{2eI_{c}}(1-I/I_{c})^{-1/2}$. This relation conforms with
the identity between the plasma frequency $\omega _p/2\pi$ and the circuit parameters
of the junction: $\omega _ {p}=(L_{J}C_{J})^{-1/2}$.

\begin{figure}
\includegraphics[width=6cm]{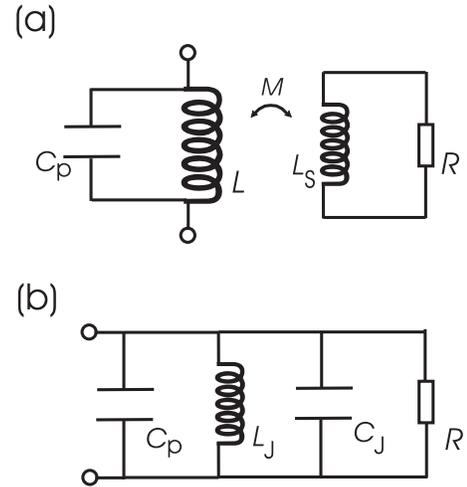}
\caption{\label{model}Circuit models of the two measuring devices: in (a) we model the
SQUID by an inductor with inductance $L_{s}$, in (b) the Josephson junction is
represented by a parallel LRC circuit.}
\end{figure}

The equivalent circuits corresponding to {\bf A} and {\bf B} are shown in
Fig.~\ref{model}a and  Fig.~\ref{model}b, respectively. In (a) we assume the SQUID to
be a pure inductor of inductance $L_{  s}$. In both (a) and (b) $C_{  p}$ represents
the total capacitance across the pump circuit, including half of the ground
capacitance $C_{0}$ of the (symmetric) pump loop. We assume that the gate capacitances
of the pump are small and can be neglected. In (b) the current biased measuring
junction is replaced by its RCSJ equivalent.

We do not consider in detail the exact electromagnetic environment. But, as will be
obvious, the value of the resistance in shunting the SQUID in {\bf A} or the junction
in {\bf B} does not influence our general conclusion. Furthermore, the LC-type lines
in the environment, not included in the present discussion, further help to decouple
the dissipative environment from the measured circuit. In {\bf A} we can assume that
the parallel amplifier impedance and in {\bf B} the dissipative (real) part of the
environment are included in $R$.

A few words on the current resolution $\delta I$ of the two circuits are in order. In
an optimum measurement by circuit {\bf A}, $\delta I$ is quantum limited to $\delta I
/\sqrt{\Delta f}\sim \sqrt{4\pi\hbar/L}$, where $\Delta f$ is the bandwidth of the
measurement. This holds for perfect coupling to the SQUID. Inserting $L=100$ nH we
obtain $\delta I /\sqrt{\Delta f}\sim 120$ fA and for the bandwidth $\Delta f=10$ kHz
we have $\delta I \sim 12$ pA. For faster measurement one needs higher input
inductance: to achieve 10 pA resolution at 1 MHz, $L$ should exceed 10 $\mu$H, which
starts to influence the pump operation itself, as will be discussed below. The current
resolution of circuit {\bf B} depends on optimizing its performance through the
measuring junction characteristics. The ultimate limit of resolution can be reached
when the junction is in the macroscopic quantum tunneling (MQT) limit, where escape
to a finite voltage state is via tunneling through the washboard potential and
thermal escape over the barrier is prohibited due to low temperature.  Using a
trapezoidal current pulse of height $I$ and duration $\Delta t$ we then have the
probability $P=1-\exp(-\Gamma \Delta t)$ to escape from the zero voltage state,
$\Gamma$ is the MQT escape rate. Since  $\Gamma$ depends on the bias current $I$, we
can use $P$ as a measure of the total current through the junction: ideally this
current is the sum of the bias current and the current to be measured, as suggested by
Vion {\em et al.}~\cite{vion}, whereby a figure of merit of the measurement is $\delta
I = \delta P/(\partial P/\partial I)$ giving the smallest resolvable current. Here,
$\delta P$ is the resolution in reading $P$, which for typical averaging is of the
order of 0.01. By a straightforward analysis we find out that there is a trade-off of
junction parameters in the sense that by reducing $\delta I$ (by increasing the ratio
$C_{J}/I_{c}$) we lower at the same rate the cross-over temperature $T_0$ from
thermally activated escape to MQT. Using this we can find an approximate answer
$\partial P/\partial I \sim (\hbar/2e)/(k_{B}T_0)$. This, in turn, with $\delta P =
0.01$, and by setting $T_0 = 30$ mK, a realistic electron temperature in an
experiment, yields $\delta I \sim 10$ pA. The bandwidth of this measurement depends on
various technical parameters in the setup and we will not discuss it here. The
comparison of the two measurements is limited by the fact that they present
fundamentally different approaches. Circuit $\bf A$ aims at a continuous monitoring of
the current at a certain bandwidth. Circuit $\bf B$ in turn tries to grab information
of the CPP current in a single shot, or by successive current pulses into the escape
junction. Both of them can achieve the requested current resolution, but the
suitability of the two depends on how they interfere with the measured CPP and whether
one can initialize the CPP controllably. One of the advantages of circuit $\bf B$ is
that the phase fluctuations are smaller owing to the smaller inductance of this
detector.

In the presence of the measuring circuit the bias phase difference $\varphi$ acquires
its own dynamics. As a result, the phase will contain a fluctuating part, $\varphi(t)
= \varphi_0 + \phi(t)$, and this will in general modify the pumped charge. Before
studying this modification in detail for the set-ups of Fig.~\ref{measure}, we analyze
the time-dependence of phase fluctuations induced by the measuring circuits.
Fluctuation-dissipation theorem implies that the variance of the fluctuations obeys
the relation
\begin{equation} \label{eq5}
\langle \phi^2(t)\rangle = 8 \int _{0}^{\infty}\frac {d\omega}{\omega} \frac
{{\Re\mbox{e}} Z(\omega)}{R_K} \coth(\frac{\hbar \omega}{2k_{B}T})[1-\cos(\omega t)].
\end{equation}
In quantifying the phase fluctuations, we thus need to know the real (dissipative)
part of the impedance seen by the measured circuit, ${\Re \mbox{e}} Z(\omega)$, where
$\omega$ is the (angular) frequency.

\noindent {\em Circuit {\bf A}} --- In this case  we obtain
\begin{eqnarray}
&&{\Re \mbox{e}} Z_A(\omega) = \nonumber \\
&&\frac {M^2R\omega^2} {R^2(1-LC_{ p}\omega^2)^2+\omega^2L_{s}^2[1-(1-\gamma)LC_{
p}\omega^2]^2}, \label{ZA}
\end{eqnarray}
with $\gamma=M^2/(L_{s}L)$. For typical experimental values this represents a narrow
resonance peak at $\omega \simeq \omega _{LC}$, where $\omega _{LC}\equiv
(LC_{p})^{-1/2}$ is the natural LC resonance (angular) frequency of the CPP circuit
alone. We first study the limit $\gamma \ll 1$, {\em i.e.}, weak coupling of the
measuring circuit. The circuit can then behave in two different ways depending on
whether the parameter $\alpha^2 \equiv (\omega_{LC}L_s/R)^2$ is either $\ll 1$ or $\gg
1$. In the first limit the resonant frequency is $\omega _{LC}$ and the width of the
peak is $\delta \omega /\omega _{LC} \simeq (M/L)^2 \sqrt{L/C_p}/(2R) = \gamma
\alpha/2 \ll 1$. The product of the width and height of the resonance peak is ${\Re
\mbox{e}} Z(\omega)_{ max} \delta \omega \simeq \omega _{LC}\sqrt{L/C_{p}}$,
independent of $R$. In the second limit, $\alpha^2 \gg 1$, which corresponds to most
experimental conditions, the measuring circuit decreases the effective inductance of
the resonant circuit and the resonant frequency (maximum of ${\Re \mbox{e}}
Z_A(\omega)$) is shifted very slightly upwards to $\omega_{LC}/\sqrt{1-\gamma}$. The
width of the peak is now $\delta \omega /\omega _{LC} \simeq
(M/L_{s})^2R/(2\sqrt{L/C_{p}})=\gamma/(2\alpha) \ll 1$. The product of the width and
height is again ${\Re \mbox{e}} Z(\omega)_{  max} \delta \omega \simeq \omega
_{LC}\sqrt{L/C_{p}}$, independent of $R$.

From Eqs.~(\ref{eq5}) and (\ref{ZA}) we see that in both limits the phase fluctuations
do not diverge with time. Instead we have initial oscillations which dephase and
asymptotically the fluctuations tend to a constant expectation value. This behavior
is not so unexpected since the measured circuit forms essentially a SQUID loop, where
the average voltage over the junctions vanishes. The decay time of coherent phase
oscillations turns out to be $t_{dec,A} = LC_pR/(\gamma L_s)$ when $\alpha^2 \ll 1$
and $t_{dec,A} = L_{s}/(\gamma R)$ in the opposite limit. We obtain the following
expressions for the variance in the short and long time limits, respectively,
\begin{eqnarray} \label{eq7}
&&\langle \phi^2(t)\rangle \simeq 4\pi (\frac{\sqrt{L/C_{p}}}{R_K})
\coth(\frac{\hbar\omega _{LC}}{2k_{B}T})\nonumber \\
&&\times \left\{
\begin{array}{ll}
[1-\cos (\omega _{LC}t)] &\mbox{ if } t \ll t_{dec,A}\\
1 &\mbox{ if } t \gg t_{dec,A}
\end{array}
\right. \;  \; .
\end{eqnarray}
In the opposite limit of strong coupling of the measuring circuit, $\gamma \rightarrow
1$, we essentially approach circuit {\bf B} to be discussed below.

\noindent {\em Circuit {\bf B}} --- In this case we have
\begin{equation}
{\Re \mbox{e}} Z_B(\omega) = \frac {L_J^2R\omega^2} {R^2(1-L_{J}C_{
J}^*\omega^2)^2+\omega^2 L_J^2}, \label{ZB}
\end{equation}
with $C_{J}^*\equiv C_{J} +C_{ p}$. In this case the low damping limit corresponds to
$\delta \omega /\omega _{LC} \simeq \sqrt{L_{J}/C_{J}^*}/R \ll 1$, where $\omega
_{LC}\equiv (L_{J}C_{J}^*)^{-1/2}$ is again the LC resonance (angular) frequency of
the CPP circuit now with the measuring junction. In this limit the maximum of the real
impedance is ${\Re \mbox{e}} Z_B(\omega)_{  max} \simeq R$ and the product of height
and width is ${\Re \mbox{e}} Z_B(\omega)_{  max}\delta \omega \simeq \omega _{LC}
\sqrt{L_{J}/C_{J}^*}$, again independent of $R$.

The coherence of oscillations in the underdamped case, $\delta \omega /\omega _{LC}
\ll 1$ dies in a time $ t_{dec,B} = 2 RC_{J}^* $, and as a result, the asymptotics of
the variance for short and long times are given by
\begin{eqnarray} \label{asympB}
&&\langle \phi^2(t)\rangle \simeq 4\pi (\frac{\sqrt{L/C_{J}^*}}{R_K})
\coth(\frac{\hbar\omega
_{LC}}{2k_{B}T}) \nonumber \\
&&\times \left\{
\begin{array}{ll}
[1-\cos (\omega _{LC}t)] &\mbox{ if } t \ll t_{dec,B}\\
1 &\mbox{ if } t \gg t_{dec,B}
\end{array}
\right. \;\; .
\end{eqnarray}

It is interesting to see how the resonance behavior characteristic for circuits {\bf
A} and {\bf B} transforms into an essentially diffusive one when the quality factor of
the resonance in the dissipative LC circuit gets lower. Let us consider circuit {\bf
B} as the example. In the overdamped case, $\delta \omega /\omega _{LC} \gg 1$, the
phase fluctuations are diffusive when $t \ll t_{dec}$. In particular, when $k_{B}T \gg
\hbar \omega _{LC}$, we have
\begin{equation} \label{eq12}
\langle \phi^2(t)\rangle \simeq 8\pi k_{B}T Rt/ (\hbar R_K),
\end{equation}
like in a purely dissipative environment~\cite{prb01}. At $t \sim \delta \omega
/\omega _{LC}^2$ the phase fluctuations level off at the value given by the long-time,
high-temperature limit in Eq.~(\ref{asympB}),
\begin{equation} \label{eq13}
\langle \phi^2(t)\rangle \simeq 8\pi k_{B}T L_{ J}/ (\hbar R_K).
\end{equation}
In the limit $\delta \omega/\omega_{LC} \gg 1$ the circuit attains features on one
hand of an inductance limited phase diffusion measurement and on the other hand of a
dissipatively dephasing environment.

\begin{figure}
\includegraphics[width=8.8cm]{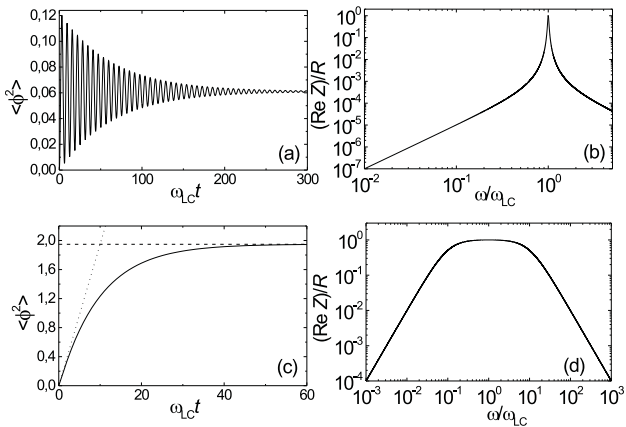}
\caption{\label{results}(a) Numerically calculated $\langle \phi^2\rangle$ for circuit
{\bf B}. In (b) we show ${\Re \mbox{e}} Z(\omega)$. The parameter values for these
graphs are: $L_{J} =$ 0.1 nH, $C_{J}^* =$ 0.1 pF, $R= $ 1 k$\Omega$, $T=$ 50 mK,
whereby $\delta \omega/\omega _{LC} = $ 0.03162 and $\omega _{LC}=$ 3.162$\cdot
10^{11}$ s$^{-1}$. (c) and (d): demonstration of the influence of the overdamped
inductive environment. For both graphs we used the following parameters: $\delta
\omega/\omega _{LC}=$ 10, $\hbar\omega _{LC}/k_{B}T=$ 0.1, $R=$ 5 $\Omega$. The
predictions of Eqs. ($\ref{eq12}$) and ($\ref{eq13}$) are shown by the dotted and the
dashed lines, respectively, and they intersect at $\omega _{LC}t=\delta \omega/\omega
_{LC}$.}
\end{figure}

Illustrations of the results of this section for model circuit {\bf B} are shown in
Fig.~\ref{results}. In Fig.~\ref{results}a we see the results of numerically
calculated $\langle \phi^2(t)\rangle$ for typical parameter values, and
Fig.~\ref{results}b shows ${\Re \mbox{e}} Z(\omega)$ with the same parameter values.
Fig.~\ref{results}c demonstrates the crossover suggested by Eqs.~(\ref{eq12}) and
(\ref{eq13}) in the dissipative case; the corresponding ${\Re \mbox{e}} Z(\omega)$ is
shown in Fig.~\ref{results}d.

The analysis of the variance of phase fluctuations performed in this section indicates
that the measuring circuits {\bf A} and {\bf B}, unlike a purely resistive
environment, will not lead to a complete suppression of coherence of the CPP: the
phase diffusion is essentially limited by the inductance. In the next section, we will
analyze the influence of these fluctuations on the amount of (coherently) transferred
charge through the pump. As we will see, fluctuations renormalize not only parameters
of the pump, but may also induce (coherent) higher order charge transfer, like
co-tunneling. Both effects lead to additional corrections to the pumped charge.

\section{Influence of the measuring circuit on the transferred charge}
\label{influence}

A convenient way to discuss the influence of the measuring apparatus on the charge
transferred during one pumping cycle is to formulate the problem in the framework of
the so-called effective action approach. This approach enables one to obtain the
partition function of the pump together with its measuring environment as a path
integral. For the case of an inductive environment, the degrees of freedom of the
environment can be integrated out approximately, and we obtain an effective action
that describes the low-energy behavior of the pump, in the presence of the measuring
circuit. From this effective action, we will obtain an effective Hamiltonian for the
pump, which is essentially the Hamiltonian~(\ref{ham}) up to two modifications: (i)
the Josephson energy $E_J$ in~(\ref{HJ}) should be replaced by a renormalized value
$E_{J,\mathrm{eff}}$ and (ii) a term $\hat{H}_\mathrm{ind}$ should be added, which
accounts for correlated tunneling events induced by the measuring circuit. Both
modifications will lead to corrections to the result~(\ref{cohcor}).

\subsection{Effective action}
\label{action} We start our analysis by considering the total equilibrium partition
function of the pump together with the measuring circuit,
$$
Z_{  tot} = \int {\cal D}\phi _{1} {\cal D} \phi_{2} {\cal D} \phi e^{-S},
$$
where the Euclidean action $S$ of the system is given by $ S = S_{C} + S_{J} + S_{
bath} $. Here, $S_{C}$ and $S_{J}$ are the actions associated to the Hamiltonians
$\hat{H}_C$ and $\hat{H}_J$ defined in Eqs.~(\ref{HC}) and (\ref{HJ}), respectively.
Specifically, the action $S_J$ is given by
$$
S_J = E_J  \sum \limits _{i=1} ^3 \int \limits _0 ^\beta d\tau \cos[\delta \phi(\tau)
_i + \phi(\tau)/3] ,
$$
where we introduced the phase differences across each junction in the CPP: $\delta
\phi _1 = \varphi_0/3 + \phi _1$, $\delta \phi _2 = \varphi_0/3 + \phi _2 - \phi _1$,
and $\delta \phi _3 = \varphi_0/3 -\phi _2$. The third term, $S_{  bath}$, is the
action describing the measuring circuit. Due to this contribution, the phase bias
$\varphi$ has acquired a dynamical part, $\phi(\tau)$, such that $\varphi(\tau) =
\varphi_0 + \phi(\tau)$. The action $S_{  bath}$ can be written as
\begin{equation}
S_{  bath} = \frac{1}{2} \int \limits _0 ^\beta \int \limits _0 ^\beta d\tau d\tau'
\phi(\tau) D_{\phi}(\tau -\tau') \phi(\tau ') ,
\end{equation}
where $\beta = 1/T$ (in this subsection we use units such that $\hbar = k_B = 1$). The
kernel $D_{\phi}(\tau)$ is related to the impedance of the measuring
circuit~\cite{leggett}, $D_{\phi}(\tau) = T \sum _{n} D_\phi (i\omega _n) e^{-i\omega
_n \tau}$, where
\begin{equation}
D_{\phi}(i\omega _n) = \frac{1}{4e^2} \frac{|\omega_{n}|}{Z(i|\omega_{n}|)}  .
\end{equation}
Here, $\omega _n$ is a bosonic Matsubara frequency, $\omega _n = 2 \pi n T$. It is
possible to get an effective action depending on the phases $\phi_1(\tau)$ and $\phi
_2(\tau)$ of the islands only. We write the total partition function as
$$ Z_{  tot} \simeq \int {\cal
D}\phi _{1} {\cal D} \phi_{2}  e^{-S_{\mathrm{eff}}}.
$$
To second order in $E_J$, $S_{\mathrm{eff}}$ is found by a re-exponentiation of the
averages $\langle S_J\rangle$ and $\langle S_J^2\rangle $ over the action $S_{bath}$.
For the average of $S_J$ we find
\begin{equation}
\langle S_{J} \rangle = E_{J,\mathrm{eff}} \sum \limits _{i=1}^{3}\int \limits _{0}
^{\beta} d\tau \cos \delta \phi _{i}(\tau), \label{SJ1}
\end{equation}
where we introduced the renormalized Josephson coupling
\begin{equation}
E_{J,\mathrm{eff}} = E_{J} e^{- G_{\phi}(\tau = 0)/18}, \label{EJeff}
\end{equation}
with the phase-phase correlation function
\begin{equation}
G_{\phi}(\tau ) = T \sum \limits_{n} D_{\phi}^{-1}(\omega _{n}) e ^{-i \omega _n \tau}
. \label{Gphi}
\end{equation}
For the average of $S_J ^2$ we obtain
\begin{widetext}
\begin{equation}
\langle S_{J}^{2} \rangle = \frac{E_{J, \mathrm{eff}}^{2}}{2} \sum \limits
_{i,j=1}^{3}  \int \limits _{0} ^{\beta} d\tau_{1} d\tau_{2} \{ \cos \delta
\phi_{i}(\tau _{1}) \cos \delta \phi_{j}(\tau _{2 }) \sum  \limits _{\sigma=\pm
1}e^{\sigma G_{\phi}(\tau _{1} - \tau _{2})/9} +  \sin \delta \phi_{i}(\tau _{1}) \sin
\delta \phi_{j}(\tau _{2}) \sum  \limits _{\sigma=\pm 1} \sigma e^{\delta
G_{\phi}(\tau _{1} - \tau _{2})/9} \} . \label{SJ2}
\end{equation}
\end{widetext}
We see that at this order an interaction term appears, whose kernel depends on the
correlation function $G_{\phi}(\tau)$. As already discussed in the previous section,
fluctuations in the bias phase are bound for the particular measuring environments we
are considering. In the limit of small phase fluctuations, the correlations $G_\phi
\sim \langle \phi(\tau) \phi \rangle$ remain small, allowing us to expand the
exponentials in~(\ref{SJ2}) with respect to $G_\phi$. Therefore we can write
\begin{eqnarray}
\langle S_{J}^{2} \rangle = E_{J, \mathrm{eff}}^{2}\sum \limits _{i,j=1}^{3}  \int
\limits _{0} ^{\beta} d\tau_{1} d\tau_{2}
[ \cos \delta \phi_{i}(\tau _{1}) \cos \delta \phi_{j}(\tau _{2 }) \nonumber \\
+ \frac{1}{9}\sin \delta \phi_{i}(\tau _{1}) G_{\phi}(\tau _{1} - \tau _{2}) \sin
\delta \phi_{j}(\tau _{2})] . \label{SJ2simpl}
\end{eqnarray}

We are now in a position to obtain the effective action $S_{\mathrm{eff}}$,
$$
S_{\mathrm{eff}} = S_C + S_{J,\mathrm{eff}} + S_{\mathrm {ind}} \;\; ,
$$
where
\begin{equation}
S_{J,\mathrm{eff}}
 =
E_{J, \mathrm{eff}} \sum \limits _{i=1}^{3}\int \limits _{0} ^{\beta} d\tau \cos
\delta \phi _{i}(\tau) \label{SJeff}
\end{equation}
and
\begin{eqnarray}
&& S_{\mathrm{ind}} =
 -\frac{E_{J, \mathrm{eff}}^{2}}{9} \sum
\limits _{i,j=1}^{3}  \int \limits _{0}
^{\beta} d\tau_{1} d\tau_{2} \times \nonumber \\
&&\sin \delta \phi_{i}(\tau _{1}) G_{\phi}(\tau _{1} - \tau _{2})\sin \delta
\phi_{j}(\tau _{2}) . \label{Sind}
\end{eqnarray}
The influence of the measuring circuit is indeed two-fold: (i) In the effective
action, the bare Josephson coupling energy $E_J$ has been replaced by an effective,
renormalized value, $E_{J, \mathrm{eff}}$; (ii) The measuring circuit correlates
tunneling events at different junctions, as is seen from Eq.~(\ref{Sind}). Physically,
a tunneling event occurring at time $\tau _1$ in junction $i$ virtually excites the
measuring environment, which, upon relaxation, causes a second tunneling event in
junction $j$ at time $\tau _2$. In other words, the environment induces higher order
tunneling events in the pump of the co-tunneling type. Clearly, both (i) and (ii) will
modify the charge transferred through the pump, as we will discuss in more detail
below.

\subsection{Renormalized Josephson coupling}
\label{EJrenorm}

In order to evaluate the renormalized Josephson coupling energy $E_{J,\mathrm{eff}}$
as given by Eq.~(\ref{EJeff}), we need to calculate the correlation function
$G_{\phi}$, Eq.~(\ref{Gphi}), at time $\tau = 0$. This implies calculating the sum
over all Matsubara frequencies of the quantity $Z(i|\omega_{n}|)/|\omega_{n}|$, which
involves the total, imaginary-time response of the circuit, $Z(i|\omega_{n}|)$. For
the circuits {\bf A} and {\bf B} of interest here we find
\begin{eqnarray}
Z_A(i|\omega|)/|\omega|= \nonumber  \\
\frac{RL+(1-\gamma)L_sL|\omega|} {R(1+LC_p |\omega|^2) +L_s |\omega|[1 + (1-\gamma)
LC_{  p}|\omega|^2]} \label{ZAim}
\end{eqnarray}
and
\begin{equation}
Z_B(i|\omega|)/|\omega|= \frac{R L_{  J}} {R(1+L_J C^*_J |\omega|^2) + |\omega|L_J},
\label{ZBim}
\end{equation}
respectively. Note that, upon analytic continuation $i\omega_n \to \omega + i0$, the
real part of $Z_{A,B}(i|\omega _n|)$ coincides with the results~(\ref{ZA}) and
(\ref{ZB}), respectively.

From a closer inspection of Eqs.~(\ref{ZAim}) and (\ref{ZBim}) we see that we can
obtain $Z_B$ from $Z_A$ upon taking the limit $\gamma \to 1$ in (\ref{ZAim}), thereby
replacing  $L$ and $L_s$ by $L_J$. We will, therefore, focus on circuit {\bf A} in
what follows; the results obtained can be used to analyze circuit {\bf B} after taking
proper limits.

\begin{figure}
\includegraphics[width=7cm]{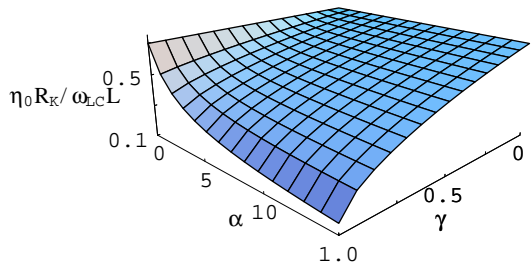}
\caption{Renormalization parameter $\eta_0$ for circuit {\bf B} in units
$\omega_{LC}L/ R_K$ as a function of $\alpha$ and $\gamma$.} \label{fig3D}
\end{figure}

In the zero temperature limit $T \to 0$, the sum over $\omega _n$ in Eq.~(\ref{Gphi})
can be replaced by an integral. In this limit, the renormalized Josephson coupling
energy can be written as $E_{J,\mathrm{eff}} = E_J e^{- \eta_0}$, where
$$
\eta_0 = \frac{4\omega _{LC} L}{9 R_K} \int \limits _0 ^\infty dx \frac{1 + \alpha
(1-\gamma) x}{1+ x^2 + \alpha x[1 + (1-\gamma)x^2]} .
$$
We introduced the dimensionless integration variable $x=\omega/\omega_{LC}$ and used
the parameter $\alpha = \omega _{LC }L_s/R$ introduced in Section~\ref{coupling}. Note
that the integral is always finite: the integrand is well-behaved, both for small
values of $x$, reflecting the fact that low-frequency phase fluctuations are limited
by the presence of an inductance in the circuit, and for large values of $x$,
reflecting the natural high frequency cut-off of the fluctuations provided by the
presence of capacitances. In Fig.~\ref{fig3D}, we plotted $\eta_0$ as a function of
the dimensionless quantities $\alpha$ and $\gamma$. Let us use some realistic
parameter values to estimate the effect. For circuit {\bf A}, $\alpha \sim 1$ and
$\gamma \ll 1$ ; hence we have $\eta _0 \alt 0.1$ and $E_{J,\mathrm{eff}} \agt 0.9
E_J$. For circuit {\bf B}, $\alpha \sim 10^{-2}$ and $\gamma \equiv 1$ ; hence we have
$\eta _0 \alt 10^{-3}$ and $E_{J,\mathrm{eff}} \simeq E_J$. We conclude that the
renormalization of $E_J$ is always weak. This result confirms the analysis of
Section~\ref{coupling}: for small values of $\alpha$ or $\gamma$ the parameter $\eta
_0$ indeed equals the asymptotic values for $\langle \phi ^2 \rangle /18$, see
Eqs.~(\ref{eq7}) and (\ref{asympB}). Thus, for suitably chosen parameters, the
measuring circuit {\bf A} (as well as {\bf B}) will not suppress the coherent coupling
between various parts of the pump.

The results presented so far are valid at zero temperature, $T=0$, and in the absence
of a gate modulation, $f=0$. Finite temperature introduces a correction to $\eta _0$
due to the summation over Matsubara frequencies in Eq.~(\ref{EJeff}). As long as the
inductance $L$ of the pump is small compared to the inductance $\hbar/(2e I_J)$ of the
Josephson junctions, this correction can be neglected at temperatures of interest here
$k_BT \ll E_J$. The effect of a finite frequency $f$ can also be ignored at this
point, since we are working in the leading order in the adiabatic approximation.

The above results, obtained for an inductive environment, should be contrasted with
the effect of a purely resistive measuring circuit, characterized by a resistance
$R_0$. In this case, the low-frequency impedance is constant, $Z \simeq R_0$, and the
sum over Matsubara frequencies in Eq.~(\ref{Gphi}) diverges logarithmically. This
reflects the fact that the low-frequency fluctuations of the phase are unbound. As a
result, the Josephson coupling will be significantly renormalized. Indeed, in the case
of a very resistive environment, characterized by a large dimensionless resistance
ratio $4R_0/9R_K \gg 1$, the renormalized Josephson coupling at $T=0$ is given by
$E_{J,\mathrm{eff}} \sim E_J (f R_0 C_p)^{4R_0/9R_K}$, which depends explicitly on the
operating frequency $f$ of the pump. This means that the Josephson energy is
completely suppressed in the absence of a gate modulation, $f = 0$. For finite values
of $f$, the pump will be characterized by a finite Josephson coupling energy; however,
its value will be small, $E_{J,\mathrm{eff}}^2/E_C \ll \hbar f$, for any reasonable
operating frequency. We thus conclude that it is practically impossible to operate the
pump in the adiabatic regime if $R_0$ is large. In the opposite limit of a weakly
resistive environment, $4R_0/9R_K \ll 1$, one finds $E_{J,\mathrm{eff}} \sim E_J (E_J
R_0C_p/\hbar)^{4R_0/9R_K} \alt E_J$: the renormalization of the Josephson coupling is
limited. We conclude that, at least in principle, the device can be operated
coherently in the adiabatic regime in the presence of weak dissipation, as long as
$R_0$ is sufficiently small.

\subsection{Calculation of the transferred charge}
\label{charge}

We finally turn to the actual calculation of the charge transferred per cycle. In
order to use Eq.~(\ref{pumpchar}), which involves instantaneous eigenstates of a
Hamiltonian, it is convenient to transform the effective action $S_{\mathrm{eff}}$ into
an effective Hamiltonian, $ \hat{H}_{\mathrm{eff}} = \hat{H}_{C} + \hat{H}_{J,
\mathrm{eff}} + \hat{H}_{\mathrm{ind}} $. With the help of Eq.~(\ref{SJeff}) we see
that $\hat{H}_{J, \mathrm{eff}}$ can be obtained directly from~(\ref{HJ}), replacing
$E_J$ by the renormalized value $E_{J,\mathrm{eff}}$. The Hamiltonian
$\hat{H}_\mathrm{ind}$ is obtained from $S_\mathrm{ind}$, which depends on the
correlation function $G_\phi$. The fact that $G_\phi$ is in general non-local in time
complicates matters. However, for the measuring circuit {\bf A}, the dynamics at low
frequencies is dominated by the inductance $L$ (for circuit {\bf B} it is $L_J$).
Provided that $L$ is small, such that $L \ll \hbar/(2eI_J)$, the relevant correlations
are local in time and $G_\phi(\tau) \simeq (8 \pi L/R_K) \delta(\tau)$. In other
words, the two correlated tunneling events induced by the measuring environment are
instantaneous on the slow time scale characterizing the junction dynamics.
Substituting the above local form for the correlator $G_\phi$ into (\ref{Sind}) and
performing one of the integrations over imaginary time we see that the resulting
$S_\mathrm{ind}$ can be transformed into
\begin{equation}
\hat{H}_\mathrm{ind} = - \frac{E_{J,\mathrm{eff}}}{18}
\frac{E_{J,\mathrm{eff}}L}{(\hbar/2e)^2} \sum \limits _{i,j=1} ^3 \sin \delta \phi _i
\sin \delta \phi _j . \label{Hind}
\end{equation}

The calculation of $Q_P$ with~(\ref{pumpchar}) with the Hamiltonian
$\hat{H}_\mathrm{eff}$ for a triangular gating sequence is straightforward. Ignoring
$\hat{H}_\mathrm{ind}$, one obtains the result (\ref{cohcor}), with $E_J$ replaced by
its renormalized value $E_{J,\mathrm{eff}}$. If $\hat{H}_\mathrm{ind}$ is taken into
account, a correction is found due to coherent higher order tunneling events induced
by the inductive environment. Referring the reader to Appendix~\ref{appind} for
details, we only present the final result:
\begin{equation} \label{centres}
Q_{P}/2e \simeq 1- \left(9\frac{E_{J,   \mathrm{eff}}}{E_{C}} - \frac{1}{6}\frac{E_{J,
\mathrm{eff}}L}{(\hbar/2e)^2}\right)\cos \varphi_0 \;\; .
\end{equation}
Taking $E_J/E_C = 0.1$ with $E_C \sim 1$K, the renormalization of $E_J$ is negligible.
With $L=100$nH (circuit {\bf A}), the second correction
$E_{J,\mathrm{eff}}L/[6(\hbar/2e)^2] \simeq 0.2$, whereas for circuit {\bf B} this
would be $E_{J,\mathrm{eff}}L/[6(\hbar/2e)^2] \simeq 2 \cdot 10^{-4}$.

\section{Discussion}
\label{discussion} The CPP has a few properties that make it an attractive object of
investigation both theoretically and experimentally. The mechanism for deviations of
the pumped charge from the quantized value are interesting in their own right. These
include Landau-Zener (LZ) band crossing, quasiparticle current, and last but not least
the topic of the current manuscript, the coherent quantum interference in Cooper pair
transport. They all can be avoided at least in principle by using a favorable
operation mode: LZ crossing can be minimized by operating the pump at low enough
frequency, $f \ll E_J^2/(\hbar E_C)$, quasiparticle current by low temperature and
careful filtering of the sample, and coherence induced corrections by dissipative
environment of the CPP.

Quantized charge transport is being investigated largely because it may eventually
fulfill the requirements of providing a modern standard for electrical current. At the
same time it would close the metrological triangle of electrical quantities: by now
both the voltage and resistance have their quantum standards, Josephson voltage and
quantum Hall resistance, respectively, and by determining the current using $I=qf$
would test the validity of Ohm's law with the same value of Planck's constant in the
two. But there are serious problems to be solved before $I$ can be determined at
sufficient absolute accuracy, error rate being smaller than $10^{-7}$ and output
current on the nA scale. The foremost problems are either the very small current
obtained in the tunnel junction based pumps, or the errors in the number of electrons
carried in the moving quantum dots in the semiconductor pumps.

Cooper pair pump may solve these problems eventually. The current limitation can
perhaps be lifted by using higher values of $E_J$ employing Josephson junctions made
of a superconductor with higher $T_c$ than that of aluminum (1 K). An obvious
candidate is niobium, where $T_c \simeq $ 9 K, whereby yielding an almost one order of
magnitude enhancement in $E_J \propto T_c$. Fabricating small junctions using Nb has,
however, turned out to be a challenge, which has not been fully solved
yet~\cite{Kim02,Dolata02}. The second problem is how to suppress errors due to the
quantum interference in the CPP without disturbing the operation of the pump
otherwise. As an example, by inserting a highly dissipative termination to the pump,
heating becomes a problem especially at the desired higher throughput currents.
Therefore understanding the nature of this interference is of importance in optimizing
the operation of the CPP. One alternative to dissipative environment is to employ
pumps with larger number of junctions, $N$, since the interference correction is
proportional to $(E_J/E_C)^{N-2}$.

\begin{acknowledgments}
We thank O. Buisson, G. Falci, P. Hakonen, T. Heikkil\"a, J. Kivioja, Ph. Lafarge, L.
L\'evy, M. Paalanen, H. Sepp\"a, and J. Toppari for useful discussions. FH
acknowledges support from Institut Universitaire de France. JP thanks the Centre de
Recherches sur les Tr\`es Basses Temp\'eratures, where part of this work was done, for
its hospitality. We acknowledge EC for financial support through grants IST-FET-SQUBIT
and HPRN-CT-2002-00144.

\end{acknowledgments}

\appendix

\section{Derivation of Eq.~(\ref{pumpchar})}
\label{appderivation}

The Hamiltonian $\hat{H}$, Eq.~(\ref{ham}), depends explicitly on the gate-charges
$n_{x1}$ and $n_{x2}$ through the charging term $\hat{H}_C$, Eq.~(\ref{HC}). As a
result of the periodic modulation of the vector $\vec{n}_x = (n_{x1},n_{x2})$ as a
function of time, the Hamiltonian becomes time-dependent. If the modulation is
adiabatically slow, Schr\"odinger's equation $i\hbar \partial |\psi_m (t) \rangle
/\partial t = \hat{H}(t) |\psi_m (t)\rangle$ is satisfied by an adiabatic state of the
form
$$
|\psi_m (t)\rangle = e^{-i\int \limits ^t dt' E_m(t')/\hbar} e ^{i\gamma _m(t)}
|m(t)\rangle ,
$$
where $|m(t)\rangle$ is an instantaneous eigenstate of $\hat{H}(t)$ with energy
$E_m(t)$ and $\gamma_m$ is Berry's phase~\cite{Berry}.

The expectation value of any time-independent hermitian operator $\hat{O}$ in the
state $|\psi _m(t) \rangle$ is in general time-dependent and given by $\langle O (t)
\rangle = \langle \psi_m (t)| \hat{O} |\psi_m (t)\rangle = \langle m (t)| \hat{O} |m
(t)\rangle$. Specifically, we see that the dynamics of $\langle O (t) \rangle $ is
governed by the instantaneous eigenstates. Using the fact that $|\dot{m}\rangle =
\dot{\vec{n}}_x \cdot \partial_{\vec{n}_x} |m\rangle$, where $\dot{a} \equiv da/dt$,
we obtain
\begin{equation}
\langle \dot{O} \rangle = 2 \Re \mbox{e } \sum _{n \ne m} \langle m | \hat{O}
|n\rangle \langle n|\partial_{\vec{n}_x} m\rangle \cdot \dot{\vec{n}}_x . \label{dyn}
\end{equation}
Note that the term $n=m$ can be excluded from the sum in~(\ref{dyn}), as $\langle
n|\partial_{\vec{n}_x} n\rangle$ is purely imaginary. This can be seen using the
normalization condition $\langle n|n\rangle = 1$ from which we obtain
$\partial_{\vec{n}_x} \langle n|n\rangle = 0 = 2 \Re \mbox{e }{\langle
n|\partial_{\vec{n}_x} n\rangle}$.

We are interested in the total charge $Q_P$ transferred through the pump during one
cycle. Since the modulation is periodic, $Q_P$ is the total charge transferred through
any of the junctions during the cycle, {\em e.g.} the leftmost one, $Q_l$. Let us
consider the matrix elements of the current passing through the left junction,
$\langle m| \hat{I}_l|n \rangle$, see Eq.~(\ref{lcur}),
\begin{equation}
\langle m| \hat{I}_l|n\rangle \equiv \frac{i}{\hbar} \langle m|
[\hat{H},\hat{Q}_l]|n\rangle = \frac{i}{\hbar} (E_m -E_n) \langle m | \hat{Q}_l
|n\rangle . \label{interm}
\end{equation}
Applying the result~(\ref{dyn}) to the charge operator $\hat{Q}_l$ and
using~(\ref{interm}), we obtain
\begin{equation}
\langle \dot{Q}_l \rangle = 2 \hbar \Im \mbox{m } \sum _{n \ne m} \frac{\langle m |
\hat{I}_l |n\rangle}{E_m -E_n} \langle n|\partial _{\vec{n}_x} m \rangle \cdot
\dot{\vec{n}}_x .
\end{equation}
Therefore, the total charge passing through the pump in the state $|\psi _m \rangle$
is given by
\begin{eqnarray}
&&Q_P =
\int _\mathrm{cycle} dt \langle \dot{Q}_l \rangle \nonumber \\
&&= 2  \hbar \Im \mbox{m} \left[ \sum \limits _{n \ne m} \oint
\frac{(\hat{I}_l)_{mn}}{E_m - E_n} \langle n| \partial _{\vec{n}_x}m \rangle \cdot
d\vec{n}_x \right].
\end{eqnarray}
which is the result Eq.~(\ref{pumpchar}) upon setting $|m\rangle = |0\rangle$.

\section{Transferred charge for triangular gating}
\label{appcharge} In this Appendix we outline the perturbative calculation of $Q_P$
with Eq.~(\ref{pumpchar}), for a triangular gating sequence as in Fig.~\ref{gating}.
This sequence corresponds to a contour consisting of three linear segments, (1), (2),
and (3), in which respectively the gate-charge vector $\vec{n}_x$ is changed
adiabatically from (0,0) to (1,0), increasing $n_{x1}$; then from (1,0) to (0,1),
simultaneously decreasing $n_{x1}$ and increasing $n_{x2}$; finally from (1,0) to
(0,0), decreasing $n_{x2}$. In this Appendix, when evaluating $Q_P$, we will present
explicit calculations for segment (1) from (0,0) to (1,0) only, {\em i.e.} we will
calculate
\begin{equation}
Q_P^{(1)} = 2 \hbar \Im \mbox{m} \left[ \sum \limits _{n \ne 0} \int \limits _0 ^1
\frac{(\hat{I}_l)_{0n}}{E_0 - E_n} \langle n| \partial _{n_{x1}}0 \rangle dn_{x1}
\right] . \label{apppumpchar}
\end{equation}
Results for the segments (2) and (3) will be simply stated, their calculation being
essentially analogous to the one for segment (1).

Since $E_J \ll E_C$, only a limited number of charge states need to be taken into
account in the calculation. This enables one to proceed perturbatively; below we will
consider various contributions to $Q_P$ that arise in different orders of perturbation
theory.

\subsection{Incoherent contribution}
\label{appincoh} In leading order, the only relevant charge states for the segment (1)
of interest here are the eigenstates $|0,0\rangle$ and $|1,0\rangle$ of $\hat{H}_C$,
Eq.~(\ref{HC}), which are mixed coherently by $\hat{H}_J$, Eq.~(\ref{HJ}). This is
analogous to the coherent mixing of charge states in a single Cooper pair box,
see~\cite{makhlin}. The two lowest instantaneous eigenstates are therefore
\begin{eqnarray}
|0\rangle = a |0,0\rangle + e^{i\varphi _0/3} b |1,0\rangle , \label{appdecom1}\\
|1\rangle = b |0,0\rangle - e^{i\varphi _0/3} a |1,0\rangle , \label{appdecom2}
\end{eqnarray}
with an energy difference $E_1 - E_0 = E_J \sqrt{1+ \epsilon^2}$. Here $a^2 = 1 - b^2
= (1/2)(1+\epsilon/\sqrt{1+\epsilon ^2})$ such that $ab = 1/(2\sqrt{1+\epsilon ^2})$.
The dependence on the parameter $n_{x1}$ enters through $\epsilon = (2E_C/3E_J)(1 - 2
n_{x1})$. Higher states can be ignored as they are separated in energy by an amount
$\sim E_C$.

We proceed by evaluating the various terms appearing in~(\ref{apppumpchar}). In order
to obtain the matrix elements $(\hat{I}_l)_{01}$ for current through the leftmost
junction, it is convenient to use the charge representation of~(\ref{lcur}). Putting
$\varphi = \varphi _0$ we find
\begin{eqnarray}
\hat{I}_l = \frac{I_J}{2i} \sum \limits _{n_1,n_2}
(e^{i\varphi _0/3} | n_1, n_2\rangle \langle n_1+1,n_2 | \nonumber \\
-  e^{-i\varphi _0/3} |n_1+1,n_2 \rangle \langle n_1, n_2 |). \label{appcharrep}
\end{eqnarray}
A direct calculation, using the above decomposition of $|0\rangle$ and $|1\rangle$
into charge states, then yields $(\hat{I}_l)_{01} = -I_J/2i$. Similarly, the
decomposition can be used to calculate $\langle 1|\partial _{n_{x1}}0\rangle =
b\partial _{n_{x1}}a - a \partial _{n_{x1}}b$. Finally, using the equality $E_1 - E_0
= E_J/2ab $, we obtain, in leading order,
\begin{eqnarray}
Q_P^{(1)} &&\simeq
4e \int \limits_0 ^1  ab (a \partial _{n_{x1}}b - b\partial _{n_{x1}}a)dn_{x1} \nonumber \\
&&\simeq 8e \int \limits _0 ^1  a(1-a^2)da  = 2e.
\end{eqnarray}

In this order, the contributions from segments (2) and (3) are zero, and we conclude
that $Q_P = 2e$. Indeed, in the absence of higher order tunneling processes, the
current through the leftmost junction is not affected by the coherent mixing of charge
states $|1,0\rangle$ and $|0,1\rangle$ along segment (2) or of $|0,1\rangle$ and
$|0,0\rangle$ along segment (3). Below, we will include higher order tunneling which
leads to corrections to these results of order $E_J/E_C$.

\subsection{Coherent correction}
\label{appcoh}

We now take into account the charge states that $\hat{H}_J$ mixes into $|0\rangle$ and
$|1\rangle$ by second order tunneling processes. For the segment (1) of interest,
these are the charge states $|0,1\rangle$ and $|1,-1\rangle$. Straightforward second
order perturbation theory yields the correction to the states $|0\rangle$ and
$|1\rangle$,
\begin{equation}
\delta |0\rangle = c |1\rangle \mbox{ and } \delta |1\rangle = c^* |0\rangle ,
\label{app2cor}
\end{equation}
where
\begin{equation}
c = (3E_J/E_C)ab(b^2 e^{i\varphi _0} - a^2e^{-i\varphi _0}). \label{appc}
\end{equation}
The energies $E_0$ and $E_1$ are also renormalized, yielding a correction to the
energy difference
\begin{equation}
\delta(E_1 - E_0)=  (6E_J^2/E_C)ab \cos \varphi _0. \label{appencor}
\end{equation}
Note the appearance in this order of a dependence of the corrections on the bias phase
$\varphi _0$. Along segment (1), phase coherence through the entire pump is
established by coherent mixing at the leftmost junction in combination with second
order tunneling through the middle and rightmost junctions.

The matrix elements of $\hat{I}_l$ are not affected by the corrections. Using
(\ref{app2cor}) we obtain $\delta (\hat{I}_l)_{01} =c^*[(\hat{I}_l)_{00} +
(\hat{I}_l)_{11}]$, which vanishes since the instantaneous eigenstates carry no
current: $(\hat{I}_l)_{00} = (\hat{I}_l)_{11}=0$, as can be seen easily from
(\ref{appdecom1}), (\ref{appdecom2}) and (\ref{appcharrep}). However, $\delta \langle
1|\partial _{n_{x1}}0\rangle$ is non-vanishing, $ \delta \langle 1|\partial
_{n_{x1}}0\rangle =  \partial _{n_{x1}}c $. The resulting correction to $Q_P^{(1)}$
can be written as the integral
\begin{eqnarray}
&&\frac{\delta Q_P^{(1)}}{2e} = \frac{E_J}{E_C} \cos \varphi _0 \int \limits_0 ^1
dn_{x1} \{
24 (ab)^3 (b\partial _{n_{x1}}a - a \partial _{n_{x1}}b) \nonumber \\
&&- 6 ab [(b^3 - 3a^2b)\partial _{n_{x1}}a - (a^3 - 3^2a)\partial _{n_{x1}}b]\} ,
\label{appint}
\end{eqnarray}
where the first term stems from the energy renormalization~(\ref{appencor}) and the
second term from the correction $\delta \langle 1|\partial _{n_{x1}}0\rangle$,
calculated taking a derivative in~(\ref{appc}). Direct integration yields
$$
\delta Q_P ^{(1)}/2e = - 3 \frac{E_J}{E_C} \cos \varphi _0 .
$$
A similar calculation for segments (2) and (3) yields the same result. In other words,
the total correction to $Q_P$ is given by $ \delta Q_{P} = - 2e (9E_J/E_C) \cos
\varphi _0 $, in agreement with Eq.~(\ref{pumpchar}).

\subsection{Correction due to inductive coupling}
\label{appind}

We finally consider the corrections to $Q_P$ associated with the Hamiltonian
$\hat{H}_\mathrm{ind}$, Eq.~(\ref{Hind}), which depends quadratically on $E_J$. Hence,
first order perturbation theory in $\hat{H}_\mathrm{ind}$ yields a correction to $Q_P$
which is linear in $E_J$. We first find the corrections to the lowest instantaneous
eigenstates
\begin{equation}
\delta |0\rangle = d|1\rangle \mbox{ and } \delta |1\rangle = - d^*|1\rangle,
\end{equation}
where
\begin{equation}
d = \{E_{J,\mathrm{eff}}L/[18(\hbar/2e)^2]\}ab(a^2 e^{-i\varphi _0} - b^2e^{i\varphi
_0}).
\end{equation}
The energy difference is renormalized as well,
\begin{equation}
\delta(E_1 - E_0)=  - \{E_{J,\mathrm{eff}}^2L/[9(\hbar/2e)^2]\}ab \cos \varphi _0 .
\end{equation}

Comparing these results with the corresponding ones found in~\ref{appcoh},
Eqs.~(\ref{app2cor}), (\ref{appc}), and (\ref{appencor}), we conclude that the
subsequent calculation will be completely equivalent to the one performed
in~\ref{appcoh}. Indeed, the matrix elements of $\hat{I}_l$ are not affected by the
corrections, and the contribution from $\delta \langle 1|\partial _{n_{x1}}0\rangle$
is equal to $\partial _{n_{x1}} d$. As a result, the correction $\delta
Q_{P,\mathrm{ind}}^{(1)}$ can be presented as an integral similar to the one in
(\ref{appint}). Segments (2) and (3) give the same contribution and the total
correction to $Q_P$ due to $\hat{H}_\mathrm{ind}$ is therefore given by
$$
\delta Q_{P,\mathrm{ind}} = 2e \{E_{J,\mathrm{eff}}L/[6(\hbar/2e)^2]\} \cos \varphi _0
,
$$
in agreement with Eq.~(\ref{centres}). The result is perturbative, the parameter
$E_{J,\mathrm{eff}}L/(\hbar/2e)^2$ must be small compared to unity. This is in
agreement with the condition $L \ll (\hbar /2e I_J)$ stated in the main text for the
derivation of the Hamiltonian $\hat{H}_\mathrm{ind}$.

\end{document}